\documentclass[11pt,a4paper]{article}
\usepackage[utf8]{inputenc}
\usepackage[T1]{fontenc}
\usepackage{lmodern}
\usepackage{microtype}
\usepackage[margin=2.5cm]{geometry}
\usepackage{setspace}
\usepackage{parskip}
\usepackage{graphicx}
\usepackage{booktabs}
\usepackage{enumitem}
\setlist[itemize]{noitemsep}
\usepackage{multicol}

\usepackage{hyperref}
\hypersetup{colorlinks=true,linkcolor=blue,citecolor=blue,urlcolor=blue}
\usepackage{titling}
\usepackage{fancyhdr}
\usepackage{caption}
\usepackage{amsmath,amssymb}
\usepackage{aas_macros}
\usepackage{natbib}

\usepackage[printwatermark]{xwatermark}
\usepackage{xcolor}
\usepackage{graphicx}

\pretitle{\vspace*{-1cm}\begin{center}\LARGE\bfseries}
\posttitle{\par\end{center}\vskip 0.5em}
\preauthor{\begin{center}\large}
\postauthor{\end{center}}
\predate{\begin{center}\large}
\postdate{\end{center}}

\begin{document}

\begin{titlepage}
  \centering
  \vspace*{0cm}
  {\Huge\bfseries Expanding Horizons \\[6pt] \Large Transforming Astronomy in the 2040s \par}
  \vspace{0.5cm}

  {\LARGE \textbf{Detached eclipsing binary star science in the 2040s}\par}
  \vspace{1cm}

  \begin{tabular}{p{4.2cm}p{10.4cm}}
    \textbf{Scientific Categories:} & Stars: binaries: eclipsing; Stars: fundamental parameters
 \\
    \\
    \textbf{Submitting Author:} & Name: Pierre F. L.  Maxted\\
    & Affiliation: Keele University, UK\\
    & Email: p.maxted@keele.ac.uk \\
    \\
    \textbf{Contributing authors:} 
    & Dominic M. Bowman,  Newcastle University, UK \\ 
    & (dominic.bowman@ncl.ac.uk) \\

    & Thomas G. Wilson, Warwick University, UK\\
    & (thomas.g.wilson@warwick.ac.uk) \\
    
    & Sophie Rosu, University of Geneva, Switzerland \\
    & (Sophie.Rosu@unige.ch)  \\
    
    & Keivan G. Stassun, Vanderbilt University, USA \\
    & (keivan.stassun@Vanderbilt.Edu)  \\

    & Simon J. Murphy, University of Southern Queensland, Australia\\ 
    & (simon.murphy@unisq.edu.au) \\

    & Ayush Moharana, Keele University, UK \\
    & (a.moharana@keele.ac.uk) \\

    & Amaury H. M. J. Triaud, Birmingham University, UK \\
    & (a.triaud@bham.ac.uk) \\

    & Axel Hahlin, Keele University, UK \\
    & (a.j.hahlin@keele.ac.uk) \\

    & John Southworth, Keele University, UK \\
    & (taylorsouthworth@gmail.com)

  \end{tabular}

  \vspace{1cm}

  \textbf{Abstract:}

  \vspace{0.5em}
  \begin{minipage}{0.9\textwidth}
    \small
    Detached eclipsing binary stars (DEBS) are currently the best source of accurate and precise fundamental stellar parameters.
    This makes DEBS crucial targets for constraining the impact of various physical processes on stellar structure and evolution. 
    Long-period binaries are particularly interesting because their separation minimises interactions between the components. This makes long-period binaries more comparable to single stars. 
    However, the current sample of DEBS with high precision stellar parameters are dominated by short-period systems (e.g. $\sim90$\,\% of the Gaia DR3 eclipsing binaries have periods $< 5$\,days).
    Facilities capable of performing detailed studies of long-period DEBS will be essential to further improve our understanding of stellar structure and evolution. Such facilities would need to be able to obtain spectroscopic observations of more distant objects at high resolution and cadence. 2--8\,m class telescopes with \'echelle spectrographs and an ability to monitor a large sample of stars would be required.
  \end{minipage}

\end{titlepage}


\section{Introduction and Background}
\label{sec:intro}

Detached eclipsing binary stars (DEBS) are our best source of accurate and precise fundamental data for most types of star. 
With modern instrumentation and data analysis techniques, it is possible to make accurate mass and radius measurements with a precision $\approx 0.1$\% in the best cases \citep{2020MNRAS.498..332M}. 
The impact of studying DEBS is widespread within astrophysics. For example, long-period DEBS containing giant stars can be used to directly measure the distances to nearby galaxies with an accuracy of 1\% \citep{2019Natur.567..200P}.
Thanks to modern surveys, such as OGLE, Kepler, TESS, Gaia, etc. \citep{2016AcA....66..405S,2011AJ....141...83P,2022ApJS..258...16P,2023A&A...674A..16M}, there are a large number of DEBS with high-quality light curves available (see review by \citealt{2021Univ....7..369S}).
Precise parallax measurements and photometry from the Gaia mission combined with all-sky photometry from ultraviolet to infrared wavelengths now makes it possible to directly measure the effective temperature of some stars in DEBS to better than 1\% \citep{2025MNRAS.tmp.1759M}.

Improved access to \'echelle spectrographs on 2--4\,m telescopes, particularly those operating in the near-infrared, has enabled us to make direct mass measurements in DEBS with extreme flux ratios \citep[$\approx 0.1$\% at optical wavelengths;][] {2022MNRAS.513.6042M,2024MNRAS.52710921S}.
With spectral disentangling techniques applied to high-quality \'echelle spectra, we can also study the abundance patterns in the stellar photospheres and directly measure the surface magnetic fields using the Zeeman effect (e.g. \citealt{2024Sci...384..214F,2025MNRAS.tmp.1963M}).
This technique is particularly effective if an uncontaminated spectrum is observed during a total eclipse \citep{2024A&A...691A.170H}.
Moreover, the flux ratio as a function of wavelength measured from this uncontaminated spectrum can then be used to improve the precision of a $T_{\rm eff}$ measurement. Spectra taken during the ingress or egress of an eclipse can be used to observe the centre-to-limb variation in the emergent spectrum from stars for which we also have precise estimates for $T_{\rm eff}$, $\log g$ and [Fe/H].

The application of asteroseismology to stars in DEBS can yield precise constraints on the physical processes within stellar interiors, such as rotation and mixing, and also directly constrain the impact of tides, mass transfer, and improve binary stellar evolution theory (see review by \citealt{2025arXiv250908426S}).
The internal stellar density profile of stars can also be probed using apsidal motion measurements \citep{2025arXiv251026306R}.
This  is a complementary approach to asteroseismology, especially for very massive stars.
Since the majority of stars are in multiple systems \citep{2023ASPC..534..275O}, it is critical to understand the physics that shapes their formation, evolution, and interaction. 
Clues to the formation mechanism for close binary systems and the impact of tides on their dynamical evolution are also provided by measurements of the Rossiter-McLaughlin (R-M) effect \citep{2022ApJ...933..227M}.
The rate of apsidal motion in binary stars can also be used to test the predictions of General Relativity if an accurate measurement  of the spin-orbit alignment of the two stars has been measured using the R-M effect \citep{2009Natur.461..373A}.  

All of the aforementioned types of data and analysis techniques are also invaluable for studying magnetic fields and refining magneto-hydrodynamical (MHD) models of stellar atmospheres.
Magnetically-driven stellar surface processes are known to affect spectral line profiles \citep{2020A&A...633A..76C,2021MNRAS.505.1699C}. Whilst the presence and strength of some stellar features may be traced through data-driven activity indicators \citep{2021MNRAS.505..830C}, other processes (such as granulation and super-granulation) may not be so easily tracked \citep{2024MNRAS.531.4238K,2024MNRAS.527.7681L} and so detailed models are required.
For example MHD models are needed to understand and correct for turbulent motions in the photospheres of solar-type stars if we are to achieve the precision in radial velocity measurements required to characterise Earth-like planets orbiting Sun-like stars \citep{2025MNRAS.539.2248F}, as stellar activity RV signals are often orders of magnitude larger than those of true Earth-twins.
Information on the magnetic field strength in solar-type stars can also be inferred from the impact they have on the limb-darkening measured from the analysis of transits in high-quality light curves \citep{2024NatAs...8..929K,2025MNRAS.537.3943F}.
In massive stars, work is ongoing to understand how magnetic fields are a consequence of binary interaction (e.g. \citealt{2019Natur.574..211S, 2024Sci...384..214F}).

The main motivation to study DEBS is to test and improve stellar evolution models, which are essential for the characterisation of exoplanet systems and to interpret the flood of data from large-scale surveys.
This has naturally increased the demand and urgency for calibrating the various free parameters in evolution models that control physical processes (e.g.\ rotation, mixing, magnetic fields) which cannot be modelled in detail. 
For example, tides in short-period binary systems force the stars to rotate synchronously with the orbit, and in late-type stars this drives increased magnetic activity and prevents ``spin-down'' via magnetic braking in the typical way for stars like the Sun.
Since rotation is closely linked to mixing processes within a star, short-period systems are not suitable for calibrating prescriptions for diffusion and mixing processes in single-star evolution models.
However, there is currently only a small subset of the known sample of DEBS that are suitable for testing stellar models at sufficiently high precision. 
The same issue applies more so to the substellar mass regime, where there are currently only two DEBS containing a pair of brown dwarfs that have been characterised in any detail \citep{2006Natur.440..311S,  2020NatAs...4..650T}.

The boundary separating binary stars that interact and binary stars that effectively evolve as single stars is not clear.
For late-type main-sequence stars, the minimum period is at least $\approx$8\,days because most stars in systems with orbital periods shorter than this are seen to rotate synchronously with the orbit \citep{2010A&ARv..18...67T}.
The orbital periods for sub-giant and giant stars in DEBS must be much longer than this ($\approx 100$\,--\,300 days) to be suitable for testing single-star evolution models. 
On the other hand, there is a significant observational bias since catalogues of eclipsing binary systems are dominated by short-period systems. 
For example, 90\% of systems in the Gaia DR3 catalogue of eclipsing binaries have orbital periods $<5$~days \citep{2023A&A...674A..16M}.
Similarly, more than half of the systems in the DEBCat\footnote{\url{https://www.astro.keele.ac.uk/jkt/debcat/}} catalogue of DEBS with precise mass and radius estimates have orbital periods $<8$\,days.

In summary, there is a strong motivation for future facilities in the 2040s to enable detailed studies of long-period DEBS. 
Since such systems are intrinsically rare (because favourable inclinations are required) and more difficult to detect (because of observational biases), the systems of interest are typically more distant, hence fainter, than those studied to-date using 1--2\,m telescopes.
This supports the need for \'echelle spectrographs on 2--8\,m telescopes with the possibility to both monitor large sample of stars (to map out their spectroscopic orbits) and perform time-critical observations (to observe eclipses and other sources of variability, such as pulsations).

\section{Open Science Questions in the 2040s}
\label{sec:openquestions}
\begin{itemize}
\item How does binarity affect stellar evolution and where is the boundary between binary systems that evolve as single stars and those that require binary stellar evolution models?
\item What is the role of multiplicity, specifically tides, synchronisation, and circularisation, in the formation and evolution of close binary systems?
\item What constraints on interior mixing processes can be derived from detailed photospheric abundance measurements in long-period eclipsing binary stars?
\item Which prescriptions for rotation, mixing, and magnetic fields can be improved and calibrated in stellar evolution models using asteroseismology and apsidal motion?
\item How do magnetic fields affect a star's structure and evolution, and are magnetic fields synonymous with interacting binary systems among massive stars?
\end{itemize}

\section{Technology and Data Handling Requirements}
\label{sec:tech}
\begin{itemize}
\item Spectrographs capable of delivering resolution $R\approx85,000$ or better covering optical ($\approx 390$--680\,nm) and near-infrared (Y-, J-, H-band) wavelengths on a range of telescope apertures from 2-m to 8-m or more.
\item Flexible scheduling to enable observations at key orbital phases.
\item Stable instrumentation ($\pm 50$\,m/s) to enable long-term monitoring for systems (over many years) and high-precision mass measurements.
\item Fully-reduced 1D spectra with accurate corrections for scattered light, \'echelle blaze function, and sky background available immediately after observation and archived in a stable format.
\item At least one telescope with an aperture $\approx 4$-m available at any time to observe time-critical events (eclipses).
\item High efficiency and short read-out times to allow for rapid sources of variability (e.g. pulsations) to be studied at fast cadence in addition to long-period binarity.
\item Telescopes sited across a wide range of longitudes will improve the capability of the facility to observe the eclipses that can last many hours in long-period DEBS.
\end{itemize}

\begin{multicols}{2}

\setlength{\bibsep}{1pt}

\bibliographystyle{mnras}
\bibliography{refs}  

\end{multicols}

\end{document}